\documentclass[12pt]{article}
\usepackage{amsmath,amssymb,amsfonts}
\usepackage{graphicx,psfrag,epsf}
\usepackage{enumerate}
\usepackage{natbib}
\usepackage{tikz}
\usepackage{pgf}
\usepackage{setspace}
\usepackage{hyperref}

\newtheorem{thrm}{Theorem}

\def\argmax{\mathop{\rm argmax}} 
\def\argmin{\mathop{\rm argmin}} 
\def\argtru{\mathop{\rm argtru}} 

\begin{document}


\title{\bf Resolving Jeffreys--Lindley Paradox}
\author{Priyantha Wijayatunga \\
             Email: \href{mailto:priyantha.wijayatunga@gmail.com}{priyantha.wijayatunga@gmail.com}  \\
              Tel: ++46 703254262 \\
               ORCID: \url{https://orcid.org/0000-0003-1654-9148}  \\
               LinkedIn: \url{https://www.linkedin.com/in/priyantha-wijayatunga-09b8b939/} }
\date{}
\maketitle

\begin{abstract}
Jeffreys--Lindley paradox is a case where frequentist and Bayesian hypothesis testing methodologies contradict with each other. This has caused confusion among data analysts for selecting a methodology for their statistical inference tasks. Though the paradox goes back to mid 1930's so far there hasn't been a satisfactory resolution given for it.  In this paper we show that it arises mainly due to the simple fact that, in the frequentist approach, the difference between the hypothesized parameter value and the observed estimate of the parameter is assessed in terms of the standard error of the estimate, no matter what the actual numerical difference is and how small the standard error is, whereas in the Bayesian methodology it has no effect due to the definition of the Bayes factor in the context, even though such an assessment is present.  In fact, the paradox is an instance of conflict between statistical and practical significance and  a result of using a sharp null hypothesis to approximate an acceptable small range of values  for the parameter. Occurrence of type-I error  that is allowed in frequentist methodology plays important role in the paradox. Therefore, the paradox is not a conflict between two inference methodologies but an instance of not agreeing their conclusions. \\
 
\noindent
\emph{keywords:} Sharp null hypothesis, $P$-value; Bayesian; Posterior probability; 
\end{abstract}






\def\argmax{\mathop{\rm argmax}} 
\def\argmin{\mathop{\rm argmin}} 
\def\arg{\mathop{\rm arg}} 

\thispagestyle{empty}

\section{Introduction}

In statistical theory there are two major methodologies for testing a hypothesis about a parameter of a given population. They are so-called the frequentist null hypothesis testing (NHT) and the Bayesian hypothesis testing (BHT) methodologies. However, researchers have confronted with some disagreements between them. An example is so-called Jeffreys-Lindley paradox that says that for a large sample a sharp null hypothesis can be rejected by the NHT methodology even if it gets high posterior odds by the BHT methodology when it  is assigned with even a small prior probability while the alternative hypothesis is assigned with the rest of the probability, diffused over the range of parameter values that it represents. The paradox was firstly stated in \cite{JH1935} and  re-exposed in \cite{DL1957}. The reader is referred to, e.g., \cite{WL2022} for more  details on the origins of the paradox, where it is shown that the paradox was clearly discussed by Sir Harold Jeffreys whose work was largely ignored by the research community.  And it was later discussed in, among others, \cite{GS1982} and \cite{BS1987} in the mathematical statistics and in  \cite{JS2013} and \cite{CR2014} in the philosophy of science.   However, those discussions have not given any clear direction for the empirical data analyst about the opposite conclusions by the two inference methodologies.  That is, still the empirical statistical analysts are confused between the two methodologies for their statistical hypothesis testing applications. For example, in  \cite{CR2017} it is claimed that still no reasonable resolution to the paradox is given in the literature as far as statistical applications are concern. Many are in favor of BHT, e.g., in \cite{BD2019} it is shown that the NHT should not be used in comparative studies in machine learning applications. 

The idea of this paper is to explain what underpins  in the paradox and then to resolve it. Apparently the solution to the paradox, or rather how to avoid contradictory conclusions, is simpler than one would expect, especially if we consider about the substantial amount of literature written on the paradox. We show that in the context of the paradox, the contradictory conclusions are due to the use of sharp null hypothesis  as a ``good" approximation to an acceptable range of values for the parameter of interest.  Generally when we say a sharp value to an unknown quantity we often do not deny any value closer to it. However when we use a sharp value in a hypothesis, a statistical significance may arise, since  in the frequentist approach the  difference between the hypothesized value of the parameter and its observed value (the estimate) is  assessed in terms of the standard error of the estimate of the parameter, no matter what the actual numerical difference between them is and how small the standard error is whereas in the Bayesian methodology even it is there  it can be ineffective. The paradox is an instance of conflict between statistical and practical significance. Occurrence of type-1 error  that is allowed in frequentist methodology plays important role in the paradox. Therefore, the paradox is not a conflict between two inference methodologies but an instance of not agreeing their conclusions.

\section{Null hypothesis testing}

\subsection{A simple example}

Consider briefly a frequentist hypothesis test result quoted in \cite{JS2013}; out of independent $104,490,000$ Bernoulli trials, $52,263,471$ are successes and $52,226,529$ are failures, therefore the observed probability of success is $0.5001768.$  When testing if the  true value of the probability of success (the parameter $p$) is $0.5$ we get a $p$-value that is lower than the level of significance $0.01.$ Therefore, the null hypothesis is rejected at the level of significance $0.01.$ Note that the standard error of the empirical estimate of the probability of success $\hat{p}$,  that is computed by $\{\hat{p}(1-\hat{p}/)/n\}^{1/2}$ is $0.00004891394.$ It is almost equal to its maximal value (which is the value we get under the assumed null hypothesis). And the $99\%$ confidence interval for the parameter in this case is $(0.5000508, 0.5003028)$ that excludes the test value of $0.5$ with a tiny margin.

Now, for the purpose of deciding if the true probability of success that is denoted by $p$ is $0.5,$ is it necessary to do a statistical hypothesis test, whether it is frequentist or Bayesian? After all the empirical estimate from a random sample of tosses is almost the same as the test value and the standard error of the estimate  is practically zero, meaning that our uncertainty about the estimate is extremely low or rather vanishing. So, what is the purpose of doing a test under these observed circumstances? In some cases, do we need to perform hypothesis tests when the sample size increases to infinity, since in the NHT methodology  the difference between the observed estimate and the test  value is assessed in terms of the numerical value of the standard error of the estimate? Even the  actual numerical difference is small, it  can be relatively large when it is measured in terms of the standard error of the estimate that is tiny. The transformed difference is interpreted in  sense of statistical laws. This is the main reason for Jeffreys to reject the NHT methodology and develop BHT methodology (see \cite{WL2022}). Isn't it sufficient that we have an observed estimate for the unknown parameter with its standard error value? 

Note that according to statistical law, namely the Weak Law of Large Numbers, in this case, $\hat{p}$ converges in probability to $p.$ And the Central Limit Theorem says how it converges to its true value (it allows us to approximate the probability that the current estimate $\hat{p}$ being at any distance from the true parameter value $p$). Recall that, for these results it is essential that we have a sample of independent random trails. 

If, more the data that we have, more accurate the knowledge about the population parameter value, and a testing procedure is used as some confirmation, then what is the point in performing  a test in case of  more information?  Shouldn't it be that having more data depreciates the need of the testing procedure? Logically, it should. One can argue that the hypothesis tests are  not sufficient for statistical inference.  But they can be necessary, especially in small sample cases but in large sample cases. In fact, often a testing procedure uses only results of an estimation procedure to make a binary decision. It only yields how an estimate differs from hypothesized value in terms of the standard error of the estimate, i.e., how probable the estimate in the partially hypothesized world (when the variation of it is assumed to be what it has been  observed empirically). Therefore, testing is redundant or rather unnecessary if the standard error of the estimate is negligible or at least too small.

One thing we should note is that in the above sharp null hypothesis testing what we have really done is using the uncertainty of the estimate (that is measured by the standard error of it) to see how far our estimated value from the hypothesized value (with respective to it). If the absolute value of the distance is more than $1.96$ then we decide that the estimated value and the hypothetical value differ significantly at the level $5\%.$ That is, in hypothesis testing we do not test about our estimate itself but test if it is close enough to the hypothesized  value.  And we are know  that we may make corresponding mistakes or errors at most  $5\%$ of the time when they are in fact close to each other  sufficiently.  

However, in the event of our data fulfilling the required assumptions and if we are ready to accept that there are no practical differences between the values $0.499$ and $0.5,$ and the values $0.5$ and $0.501,$ i.e., the range $(0.499, 0.501)$ can be regarded as the single value $0.5$ then we should accept our null hypothesis at $0.01\%$ level of significance since the range $(0.499, 0.501)$ has an overlap with $99\%$ confidence interval. Note that we need only a partial overlap.  Recall the words of Tukey: `\emph{`It is foolish to ask `are the effects of $A$ and $B$ different?' They are (almost) always different for some decimal place."} \cite{JohnTukey1991} (p 100).

\subsection{Replication Crisis and $p$-Values}

Suppose that, unknown to us, there is no difference between the true and our hypothetical values of the parameter but in our initial experiment we saw a statistical significance. In subsequent studies we have not seen any statistical significance at the same level of significance.  That is, it has happened  a replication problem which is generally referred as replication crisis in science \cite{CS2021}.

Apart from committing a type-I error, there could be some errors in our data in the first experiment, e.g., the random sample assumption might not have fully fulfilled. However, in the long run we can be proved to be correct as long as such assumptions are fulfilled in subsequent studies! Note that Fisher's advice was to repeat the experiment several times before accepting a significant result.  If subsequent experiments also had some problems in fulfilling the required assumptions, unknown to us, then we might have seen statistical significance in them too. In this sense, it is hard to blame  the NHT methodology  alone for replication crisis, but our mechanism of data collection, etc., i.e., fulfilling the required assumptions. 

Although there exist elegant statistical methods to overcome some of the deficiencies of the data to a certain extent, e.g., in case of being aware of the data generating process, no method can be superior than having accurate and clean data. No exception for significance testing and erroneous conclusions  might be due to violation of required assumptions in the data but in the testing methodology itself. Replication problem may be  due these errors. However, in the literature, the NHT methodology as a whole  is blamed heavily for the crisis. Such acts are due to misunderstanding of the methodology.

Now let us see how observed  $p$-value can be erroneous. Here we discuss the potential uncertainty that we may neglect in calculating $p$-values. We can show  that often we may need to inflate our calculated $p$-values, especially in small or moderate size data samples. We obtain the $p$-values under some assumptions, but we are not sure if the assumptions are fulfilled. Therefore, there  is some uncertainty in them, which we often ignore. The calculated $p$-values  may be adjusted for compensating the uncertainty in our research design or in the modeling process.  

Suppose we are performing a one-sided $T$-test for the population parameter mean, since we have seen that observed sufficient statistics of the parameter is relatively larger than the assumed parameter value. Let the  sample size is $n$ and then the  test statistic $T$ has a $t$-distribution with $(n-1)$ degrees of freedom under the null hypothesis $H,$ assuming the sample of data is random. Then theoretical $p$-value of the test is the conditional probability of the event $\{T \geq t_{ob} \} $ given that $H_0$ is assumed, i.e., $p=P \{ T \geq t_{ob} \vert H_0\}, $ where $t_{ob}$ is the observed value of  $T.$  However, practically we are uncertain that our data sample is completely random. Therefore, our observed $p$-value should be written as, $p_{o}=P\{T \geq t_o,R  \vert H_0\},$ where $R$ denotes the proposition that the random sample assumption is true. In other instances, it should represent all the modeling assumptions and the assumptions about the data. Note that $p_{o}$ is the chance that two events of $\{ T \geq t_{ob}  \} $ and $ R $ are happening jointly under $H_0$. 

 \begin{eqnarray*}
  p & = &P\{T \geq t_o \vert H_0 \} = P\{T \geq t_o \vert H_0,R \}  \\
     &= & \frac{P\{ T \geq t_o, R \vert  H_0 \}}{P\{ R \vert H_o\}}=  \frac{P\{ T \geq t_o, R \vert  H_0 \}}{P\{ R \} }  \\
     & \geq &  P\{  T \geq t_o,  R \vert H_0  \} = p_{o} 
 \end{eqnarray*}
 We often assume that $P(R)=1$ even though  $P(R) \leq 1.$ Therefore, our  $p_{o}$ is often not the value that it should be! Often it can be smaller than what it should be, because, i.e.,  the applied researcher may be biased towards the alternative hypothesis when data are collected. Therefore, we tend get a significant result but often it should be otherwise. Such cases may be one of the main causes of replication problems, especially when subsequent experiments tend to fulfill the assumptions (here the random sample assumption of data).

In order to get the true $p$-value for the above test from its observed $p$-value ($p_o$-value) by inflating it, we need to estimate the probability that the modeling and other assumptions on the data are fulfilled. However, objective estimation of this probability can be harder. We avoid discussion of such matters since it is beyond the scope of this paper. Important point that we want to raise here is that appropriate handling of the uncertainty leads to reduce unnecessary problems. Appropriate use of the probability for handling uncertainty is beneficial for the purpose.

One my wonder why we may only inflate the observed $p$-value but not deflate, e.g., suppose the empirical researcher is biased towards the null hypothesis so that his/her data collection task favors it more. And mathematically it is not possible to have a deflation with above definition of the observed $p$-value. Since a $p$-value is a conditional probability where condition is that the null hypothesis is true. So, logically there is no need to reduce $p$-value from its observed value, as under the assumption of null hypothesis is true we may not try to falsify it.

\section{The Paradox}

Let us have two quotes from two researchers who addressed the paradox earlier. 

Christen P. Robert: \emph{In my opinion, Lindley’s (or Lindley-Jeffreys’s) paradox is mainly about the lack of significance of Bayes factors based on improper priors,} in   \cite{CR2013}.

D J Johnstone: \emph{More than a ``paradox" this result amounts philosophically to reductio ad absurdum of Fisher's logic for tests of significance,}  in \cite{DJJ1986}.

Sharp values for unknown quantities, population parameters in our case, are rarely used in practice. If it is done, possibly along with other restrictions,  then there can be confusions just as in case of Jeffreys-Lindley paradox, that we discuss here. In the above case, we may accept the value $0.5$ as the true probability of success (or rather a strongly reliable estimate of it)  if the observed estimate is in the closed interval, e.g., $[0.4995, 0.5005]$ that can be regarded as ``an acceptable small range of values" for the population parameter. Let us assume that anything outside this interval can be considered as the true probability not being equal to $0.5.$ If it is the case then we have to accept  the null hypothesis that the true probability of success is $0.5$ at a certain  level of significance, ideally  by redefining the numerical value of the $p$-value of the test manually (or just ignoring the test altogether). This is because there is an overlap of the $99\%$ confidence interval and the interval of acceptance $[0.4995, 0.5005].$ In fact, in this case the empirical estimate is contained in the small range of acceptance, therefore confirming the null hypothesis to a greater extent. No comparison of the $p$-value and the level of significance is needed, but the conclusion can be stated with a level of significance corresponding to the acceptable  small range of values, i.e., the level corresponding to the critical test statistics value of $3.615.$ 

For a massive sample size that is practically infinite, the estimator of the parameter has an extremely narrow probability density even though related observed  test statistic value can vary over a large range. So, before computing the test statistic value one should think about  the  probability density of the estimator (sampling distribution)---if it is meaningful to calculate probabilities from such a density. In the above case, the sampling density is almost the unit probability mass for all practical reasons! Recall that even a density over a tiny range of values can be transformed into a density over a larger range of values,  as in the case of that of the test statistic above. This is nonsense as long as our real objective is to use the observed estimate in practice but the test statistic! So, our proposal is that while working with a sharp null hypotheses using an acceptable range of values for the parameter  can prevent us from such controversies such as this paradox. 

Now let us consider the Jeffreys-Lindley paradox. As considered in \cite{BS1987}, we take a random sample $\bold{X}_n=\{X_1,...,X_n\}$ of size $n \geq 1$ from a population with the probability density $f(x \vert \theta)$ where $\theta$ is an unknown parameter with a state space $\Theta \subset \mathbb{R}.$ And we are testing the null hypothesis $H_0: \theta = \theta_0$ versus the alternative hypothesis $H_a: \theta \neq \theta_0,$ where $\theta_0 \in \Theta$ is a specific value of $\theta$ of our choice. As the authors discussed, it is rare that we are interested in a point value for an unknown parameter but a small range  of values that is realistically acceptable. But a point value is good approximation to such a small range for simplicity of computation, etc. However, if it is done so, as shown above and discussed below, sometimes paradoxical conclusions may arise.  In the NHT framework, we test the $H_0$ with the test statistic $T(\bold{X}_n),$ therefore the $p$-value of the test is $p=P\{T(\bold{X}_n) \geq \vert T(\bold{x}_n) \vert \}$ where $\bold{x}_n$  is the observed sample (observed values are denoted by lowercase).

For simplicity, let $f(x \vert \theta)$ be the normal probability density with unknown mean $\theta$ and known variance $\sigma^2.$ So, $T(\bold{X}_n)=  \sqrt{n} \vert \bar{X}_n - \theta_0 \vert / \sigma $ where $\bar{X}_n$ is the mean of the sample $\bold{X}_n$  and $p= 2(1- \Phi(t_n))$ where $\Phi$ is the standard normal cumulative distribution function and $t_n=T(\bold{x}_n)$  is the observed value of test statistic  (for observed data $\bold{x}_n).$ Assume that the observed $p$-value, $p$ is smaller than the level of significance of the test, say, $0.05,$  for the current sample which is not a large sample, therefore $H_0$ is rejected at that level. 

Now, we remind the reader what a Bayesian might do, as shown in  \cite{BS1987}. A Bayesian may assume a prior probability of, say, $\pi_0=1/2$ for $H_0,$ therefore $P(H_0)=P(H_a)=1/2.$ But, since $H_a$ is specifying a range of values for the parameter we need to spread its probability mass over that range with a probability density $g(\theta),$ say, a normal distribution with mean $\theta_0$ and variance known $\sigma_1.$ Usually $\sigma_1 >> \sigma,$ but it can be that both are equal. Then the marginal distribution of $X,$

\begin{equation*}
m(x) = f(x \vert \theta_0) \pi_0 + (1-\pi_0)m_g(x)
\end{equation*}
where
\begin{equation*}
m_g(x) = \int_{\theta \neq \theta_0} f(x \vert \theta)g(\theta)d\theta.
\end{equation*}
So, the posterior probability of $H_0$ (assuming $f(x \vert \theta_0)>0$) is
\begin{eqnarray*}
P\{H_0 \vert x\} &=& f(x \vert \theta_0) \pi_0 / m(x) \\
                        &=& \Bigg[1 + \frac{(1-\pi_0)}{\pi_0} \frac{m_g(x)}{f(x\vert \theta_0)} \Bigg]^{-1}
\end{eqnarray*}
Therefore, the \emph{posterior odds} of $H_0$ to $H_1$ is 
\begin{equation*}
\frac{P\{H_0 \vert x\}}{1- P\{H_0 \vert x\}} = \frac{\pi_0}{1-\pi_0} \times \frac{f(x \vert \theta_0)}{m_g(x)}
\end{equation*}
where $\pi_0/(1-\pi_0)$ is the \emph{prior odds} and $B_g(x)=f(x \vert \theta_0)/m_g(x)$ is the \emph{Bayes factor for $H_0$ versus $H_1.$}

For simplicity assume $\sigma_1= \sigma.$ It is shown that, for the observed data $\bold{x}$ the Bayes factor (actual odds of the hypotheses implied by the data alone) is
\begin{eqnarray*}
B_g(\bold{x})&=& \frac{f(\bar{x} \vert \theta_0)}{m_g(\bar{x})} = \frac{[2\pi \sigma^2/n]^{-1/2} exp \Big( - \frac{n}{2} (\bar{x}-\theta_0)^2/ \sigma^2 \Big)}{[2\pi \sigma^2(1+n^{-1})]^{-1/2} exp \Big( - \frac{1}{2} (\bar{x}-\theta_0)^2/ [\sigma^2(1+n^{-1})] \Big)} \\
    &=& (1+n)^{1/2} exp \Big( - \frac{1}{2} t^2_n/(1+n^{-1}) \Big)
\end{eqnarray*}

And we have
\begin{eqnarray*}
P\{ H_0 \vert \bold{x}\}= \Bigg[1 + \frac{1}{(1+n)^{1/2}} exp \Big( \frac{t^2_n}{2(1+n^{-1})} \Big) \Bigg]^{-1}
\end{eqnarray*}
It has been shown in the literature that, if $t_n$ is fixed, say, $t_n=t$ for all large $n$ (corresponding to a fixed $p$-value),  then $P\{H_0 \vert \bold{x}\} \rightarrow 1$ as $n \rightarrow \infty,$ no matter how small the $p$-value is. Note that when $n \rightarrow \infty$ the test statistic has a standard normal distribution under $H_0$, therefore for a fixed value $t_n$, its $p$-value is fixed.   Therefore, the $H_0$ is strongly favoured by the Bayesian inference method even though it is rejected by the frequentist inference method since the  $p$-value is smaller than the selected significance level always.  

This is an instance of so-called \emph{Jeffreys-Lindley paradox.} These  contradictory  conclusions leave the applied statistician and the empirical analyst  in a serious confusion. And theoretically,  this is seen as an irreconcilability of $p$-values and evidence in the data as shown in \cite{BS1987} or even the frequentist and the Bayesian inferences.

Since $ t_n =t < \infty$ as $n <  \infty,$ for any large positive integer $n$ and positive integer $m$ we have $t_n=t_{n+m};$
\begin{eqnarray*}
 \frac{\bar{x}_n-\theta_0}{\sigma/\sqrt{n}}= \frac{\bar{x}_{n+m}-\theta_0}{\sigma/\sqrt{n+m};}
 \end{eqnarray*}
This implies that 
\begin{eqnarray*}
  \frac{(\bar{x}_n-\theta_0)^2}{(\bar{x}_{n+m}-\theta_0)^2} -1 = \frac{m}{n} >0
\end{eqnarray*}
So, $m>0$ exists if 
\begin{eqnarray*}
 \frac{(\bar{x}_n-\theta_0)^2}{(\bar{x}_{n+m}-\theta_0)^2} > 1
\end{eqnarray*}
 implying that $\lim_{n\to\infty} \bar{x}_n = \theta_0.$ So, $H_0$  must be true. Therefore, in the case fixed $t_n$ for $n>0$, there happens a type-I error in the NHT methodology.   This is allowed in the methodology.  Even thought two conclusions from two hypothesis testing methodologies are opposing, there is no conflict between them, since NHT methodology allows errors, in this case type-I errors.  

And on the other hand, if the true parameter value is different from what is assumed under the null hypothesis, 
\begin{eqnarray*}
\lim_{n\to\infty} \bar{x}_n = \theta_a \neq \theta_0
\end{eqnarray*}
 for some parameter value $\theta_a$ then 
 \begin{eqnarray*}
  \lim_{n\to\infty} t_n =  \lim_{n\to\infty} \frac{\bar{x}_n-\theta_0}{\sigma/\sqrt{n}}= \infty.
  \end{eqnarray*}
  That is, $t_n$ cannot have a fixed finite value for all $n>0$. 

\section{Statistical Significance and Practical Significance}

In the frequentist inference, generally statistical significance prevails, even for tiny value of  $\vert \bar{x}_n - \theta_0 \vert$ for large samples.  For anyone who is aware of this fact, Jeffreys-Lindley paradox is not a paradox! Recall the words of John Tukey: \emph{``It is foolish to ask `are the effects of $A$ and $B$ different?' They are (almost) always different for some decimal place."} \cite{JohnTukey1991} (p 100). Also recall that it is rare that we are interested in a sharp value for the unknown parameter. We often are interested in a small range of values for it. So, it is absurd to reject a point value when it differs from the estimated value by a tiny margin if the confidence interval  has an overlap with the acceptable small range values, or at least the estimate is contained in the latter. We are doing the test using a sharp null hypothesis because it is just a good approximation for desired range of values, mainly for computational simplicity. If we accept a small range of values for unknown parameter then we can see that the paradoxical conclusions may not arise. 
 
 Let us assume that the acceptable small range of values for the parameter $\theta$ according to $H_0,$  ideally depicting so-called practical significance,  is the closed interval $[ (1 - \delta)\theta_0, (1 + \delta)\theta_0]$ where $0< \delta<1$ is small, e.g., $\delta=0.05$ for our context.  Then, if the confidence interval for $\theta$ intersects with the acceptable interval for it, then we can   accept the  $H_0$ ignoring the $p$-value of the test or defining it to be larger than the significance level. In this way, we can combine the statistical significance and  practical significance together, especially for large samples.

\section{Adjusting the null hypothesis testing procedure} 

Simply we can extend the above method of selecting a numerical value for the $p$-value to more general contexts, in order to tackle the problem of conflict between statistical and practical significances, especially for large samples.  However, it is possible only in cases where specification of the practical significance can be done. We argue that the NHT procedure should allow the empirical researcher to assign a meaningful numerical value contextually to the $p$-value by considering the acceptable small range of values for the unknown parameter. Alternatively, boundary points of the acceptable range of values can be transformed as critical values of the test statistics, i.e.,  we define the level of significance according to the sample size, especially when it is large. Such tasks should treat the problem of the controversy between the statistical and practical significances. That is, we should be able to adjust the testing procedure so that it takes into account what it really means to be a difference between the observed numerical estimate of the parameter and the hypothesised value of it, i.e., the practical significance. Practically an infinite sample is a massive sample and the sampling distribution of the estimator is often only over a tiny range of values, like in the case of Jeffreys-Lindley paradox. Therefore, in such cases performing a significance test is meaningless. But, if we perform it then we should be able to select a numerical value for the $p$-value manually overriding the value generated by the test. It may make us to accept the null hypothesis.

{}


\begin{thebibliography}{}

\bibitem[Berger and Sellke(1987)]{BS1987}
Berger, J. O. and Sellke, T. (1987). Testing a Point Null Hypothesis: The Irreconcilability of P Values and Evidence.  Journal of the American Statistical Association, \textbf{82}(397):112--122.

\bibitem[Berrar and Dubitzky(2019)]{BD2019}
Berrar, D. and Dubitzky, W.  (2019). Should significance testing be abandoned in machine learning? International Journal of Data Science Analytics, \textbf{7}:247--257. \url{https://doi.org/10.1007/s41060-018-0148-4}


\bibitem[Colling and Szűcs(2021)]{CS2021}
Colling, L.J. and Szűcs, D. (2021). Statistical Inference and the Replication Crisis. Review of Philosophy and Psychology \textbf{12}:121--147. \url{https://doi.org/10.1007/s13164-018-0421-4}


\bibitem[Cousins(2017)]{CR2017}
Cousins, R. D. (2017). The Jeffreys-Lindley paradox and discovery criteria in high energy physics, Synthese,  \textbf{194}:395--432. \url{https://doi.org/10.1007/s11229-014-0525-z}


\bibitem[Jefferys(1935)]{JH1935}
Jeffreys, H. (1935). Some tests of significance, treated by the theory of probability. Proceedings of the Cambridge Philosophy Society \textbf{31}: 203--222.

\bibitem[Johnstone(1986)]{DJJ1986}
Johnstone, D, J. (1986).  Tests of significance in theory and practice. American Statistician  \textbf{35}: 491--504.

\bibitem[Lindley(1957)]{DL1957}
Lindley, D. V. (1957). A Statistical Paradox.  Biometrika, \textbf{44}:187--192

\bibitem[Robert(2013)]{CR2013}
Robert, C. (2013) Paradoxes in Scientific Inference. Chance, 26(2), pp. 52–54. doi: 10.1080/09332480.2013.794623.

\bibitem[Robert(2014)]{CR2014}
Robert, C. P. (2014). On the Jeffreys-Lindley Paradox.  Philosophy of Science, \textbf{81}:216--232.

\bibitem[Shafer(1982)]{GS1982}
Shafer, G. (1982). Lindley's Paradox. Journal of the American Statistical Association, \textbf{77}(378):325--334.

\bibitem[Sprenger(2013)]{JS2013}
Sprenger, J. (2013). Testing a precise null hypothesis: The case of Lindley's paradox. Philosophy of Science, \textbf{80}(5):733--744.

\bibitem[Tukey(1991)]{JohnTukey1991}
Tukey, J. (1991). The philosophy of multiple comparisons. Statistical Science, \textbf{6}(1):100-116.

\bibitem[Wagenmakers and Ly(2022)]{WL2022}
Wagenmakers, E. -J. and  Ly, A. (2022). History and nature of the Jeffreys–Lindley paradox. Archive for History of Exact Sciences, \url{https://doi.org/10.1007/s00407-022-00298-3}


\end{thebibliography}
\end{document}